\documentclass[fleqn]{2023SCGE}
\usepackage[utf8]{inputenc}
\usepackage[numbers]{natbib}
\usepackage{hyperref}
\let\WriteBookmarks\relax
\def\floatpagepagefraction{1}
\def\textpagefraction{.001}

\usepackage{xcolor} 
\usepackage{colortbl} 
\usepackage[normalem]{ulem}
\usepackage{amssymb}
\usepackage{amsmath}
\usepackage{mathtools}
\usepackage{color}
\usepackage{enumitem}
\usepackage{textcomp}
\usepackage{threeparttable}
\usepackage{color}
\usepackage{epsfig}
\usepackage{color}
\usepackage{graphicx}
\usepackage{longtable}
\usepackage{subfigure}
\usepackage{soul, subfiles}
\usepackage{CJK}

\usepackage{caption2}

\newcommand{\cahk}{Ca \scriptsize{\uppercase\expandafter{\romannumeral2}} \normalsize H and K }

\begin{document}
\begin{CJK*}{UTF8}{gbsn}
\let\WriteBookmarks\relax
\def\floatpagepagefraction{1}
\def\textpagefraction{.001}
\ensubject{subject}
\ArticleType{Article}
\Year{2026}
\Month{June}
\Vol{??}
\No{??}
\DOI{??}
\ArtNo{000000}
\ReceiveDate{??}
\AcceptDate{??}
  
\title {Flare waiting time as a novel proxy of stellar magnetic activity}


\author[1]{Henggeng Han}{}
\author[1,3]{Song Wang}{{songw@bao.ac.cn}}
\author[2]{Chuanjie Zheng}{}
\author[2]{Cunshi Wang}{}
\author[2]{Xue Li}{}
\author[1,2,4]{Jifeng Liu}{}
\AuthorMark{Han H G, Wang S, Zheng C J}
\AuthorCitation{Han H G, Wang S, Zheng C J Q et al}
\thanks{Corresponding author. Email: songw@bao.ac.cn}

\address[1]{National Astronomical Observatories, Chinese Academy of Sciences, Beijing 100101, People's Republic of China}
\address[2]{School of Astronomy and Space Science, University of Chinese Academy of Sciences, Beijing 100049, People's Republic of China}
\address[3]{Institute for Frontiers in Astronomy and Astrophysics, Beijing Normal University, Beijing, 102206, People's Republic of China}
\address[4]{New Cornerstone Science Laboratory, National Astronomical Observatories, Chinese Academy of Sciences, Beijing, 100012, People's Republic of China}

\abstract{Stellar flares have long served as stellar magnetic activity tracers. The flare waiting time, defined as the interval between two consecutive flares, provides a valuable diagnostic for probing underlying mechanisms of energy storage and release in stellar atmospheres. In this work, utilizing flaring M dwarfs observed by the \emph{Kepler} satellite, we establish a simple yet effective activity proxy, i.e., median flare waiting time ($t_{\rm{w, med}}$). Our results show that the $t_{\rm{w, med}}$ can trace long-term activity levels similar to the flare rate. However, $t_{\rm{w, med}}$ corresponding to different waiting time percentiles may encode richer physical insights than flare rate. In addition, for the first time we construct a clear relation between $t_{\rm{w, med}}$ and stellar rotation period, which is quite similar to the canonical activity--rotation relation. More intriguingly, this relation exhibits a more notable supersaturation effect (i.e., below a critical rotation period, $t_{\rm{w, med}}$ begins to increase instead of keeping constant) compared to other activity proxies. The filling factor--rotation period relation favors poleward migration of active regions as the explanation for supersaturation, rather than coronal stripping. With the dramatic increase in stellar flares detected by missions like TESS and the upcoming Earth 2.0 satellite, $t_{\rm{w, med}}$ will become a powerful diagnostic for probing stellar magnetic activity and underlying physics.  
}

\keywords{Stellar activity, Stellar rotation, Main-sequence: late-type stars, Flare stars}

\PACS{97.10.Jb, 97.10.Kc, 97.10.Wn, 97.20.Jg, 97.30.Nr}

\maketitle

\begin{multicols}{2}
\section{Introduction}
\label{sec:intro}

Stellar flares are impulsive bursts of energy released by magnetic reconnection, leading to enhanced emission across multiple wavelengths \citep{2010ARA&A..48..241B}. 
Owing to the high-precision photometry from \emph{Kepler} and TESS satellites, numerous studies have compiled extensive catalogs of stellar flares, encompassing a vast number of events \citep[e.g.,][]{2016ApJ...829...23D, 2019ApJS..241...29Y, 2020ApJ...890...46T, 2022ApJ...935..143P, 2023A&A...669A..15Y, 2025ApJS..281...52W}, which enable extensive studies of flare morphology \citep{2014ApJ...797..122D, 2026MNRAS.545f2018T}, frequency distributions \citep{2012Natur.485..478M, 2013ApJS..209....5S, 2014ApJ...797..121H}, and occurrence rates \citep{2014ApJ...792...67C, 2020ApJ...905..107M}.
Flares are valuable for tracing stellar activity, probing dynamo processes, and understanding coronal heating mechanisms.
\begin{figure*}
\centering
\includegraphics[width=0.98\textwidth]{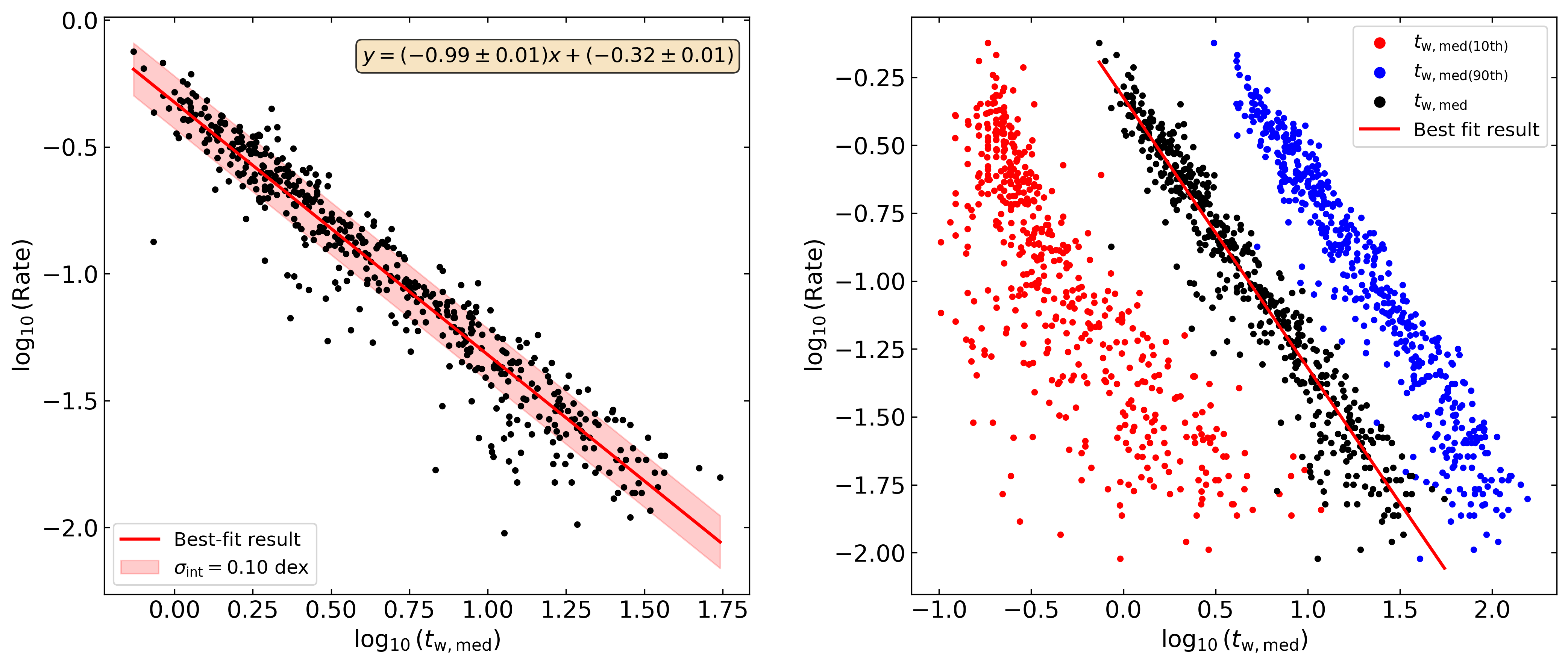}
\caption{Left panel: Relation between $t_{\rm{w, med}}$ and flare rate. Red line represents best-fit result. Shaded area represents intrinsic scatter. Right panel: Comparison between $t_{\rm{w, med}}$ corresponding to different percentiles of overall waiting time and flare rate.}
\label{picture_rate.fig}
\end{figure*}

The stellar activity--rotation relation is one of the most important observational tracers of the stellar dynamo. Historically, such relations have been constructed with diverse activity proxies, including X-ray emission \citep[e.g.,][]{2011ApJ...743...48W, 2003A&A...397..147P, 2019A&A...628A..41P, 2024A&A...684A.121F}, ultraviolet continuum flux \citep[e.g.,][]{2024ApJ...966...69L, 2024ApJ...977..138H}, chromospheric emission lines \citep[e.g.,][]{1984ApJ...279..763N, 2017ApJ...834...85N, 2023ApJS..264...12H, 2026ApJS..283...79Z}, and rotation-modulated light curve amplitude \citep[e.g.,][]{2019ApJS..244...21S, 2021ApJS..255...17S}. Meanwhile, parameters related to flares, including flare frequency distributions, normalized flare energy, and occurrence rates \citep{2017ApJ...849...36Y, 2019ApJ...871..241D, 2020ApJ...905..107M, 2021A&A...645A..42I}, have also been used to construct such a relation.
The activity--rotation relation typically exhibits a saturated regime and a decay regime. However, recent studies have revealed some fine structures. 
In the extremely fast-rotating region, activity levels decline, known as the supersaturation effect \citep{1996A&A...305..785R, 2001A&A...370..157S, 2011ApJ...743...48W, 2016A&A...589A.113A}. 
Meanwhile, within the decay regime, sub-regimes associated with core-envelope coupling process have been identified \citep{2025A&A...699A.251Y, 2026ApJ..1000..208H}. 

An underexplored flare property is the waiting time, the interval between successive flares, which tests whether flares are stochastic or causally linked. While some studies find no correlation \citep{1976ApJS...30...85L, 2014ApJ...797..121H}, others propose sympathetic triggering \citep{1995MNRAS.277..423P, 1999ApJ...515..746O}.  Yet, the potential of waiting time as an activity proxy and its role in activity--rotation relations remains largely unexplored, offering a promising new direction for dynamo diagnostics. In this work, we investigate the statistical properties of stellar flare waiting times using the flare catalog from ref. \citep{2017ApJ...849...36Y}. Our results show that median flare waiting time for individual M dwarfs can serve as a novel proxy for probing the stellar activity--rotation relation. Furthermore, we identify a clear supersaturation regime in the activity--rotation relation. We further discuss possible explanations for supersaturation, including polar updraft migration and coronal stripping.
This paper is organized as follows: in Section \ref{sec:samp} we describe our data reduction process. We present our results and discussion in Section \ref{sec:res}. A brief summary is given in Section \ref{sec:sum}. 

\section{Sample}
\label{sec:samp}

Ref. \citep{2017ApJ...849...36Y} identified 540 flaring M dwarfs in the \emph{Kepler} field, detecting approximately 103,178 stellar flares. Their catalog provides flare energies as well as the start and end times of each event. 
In this work, we define the flare waiting time as the interval between the start time of a given flare and the end time of the preceding one.

\begin{table}[H]
  \centering
  \setlength{\tabcolsep}{4pt} 
  \footnotesize
  \caption{Calculated median waiting times and filling factors}
  \label{tab:star}
  \begin{tabular}{cccccc}  
    \toprule
    KIC & $T_{\rm{eff}}$ & $P_{\rm{rot}}$ & log$_{10}(t_{\rm{w, med}})$ & $\log_{10} \bigl( \frac{A_{\mathrm{spot}}}{A_{\mathrm{star}}} \bigr)_{P_{\mathrm{rot}}}$ & $\log_{10} \bigl( \frac{A_{\mathrm{spot}}}{A_{\mathrm{star}}} \bigr)_{5\times P_{\mathrm{rot}}}$ \\%
     & (K) & (days) & & & \\ 
    
    \midrule
    892376 & 3973 & 1.53 & 0.344 & 0.0107 & 0.0197 \\
    1572802 & 3876 & 0.37 & 1.299 & 0.0871 & 0.1042 \\
    1872885 & 3847 & 1.87 & 0.269 & 0.1057 & 0.118 \\
    1873543 & 3850 & 11.55 & 0.455 & 0.0237 & 0.0301 \\
    2157356 & 3651 & 13.61 & 1.342 & 0.0251 & 0.0328 \\
    2283749 & 3694 & 1.62 & 0.33 & 0.0902 & 0.0972 \\
    2300039 & 3408 & 1.71 & 0.208 & 0.0154 & 0.0193 \\
    2424688 & 3958 & 11.1 & 1.295 & 0.0567 & 0.0702 \\
    2441562 & 3611 & 10.1 & 0.963 & 0.0321 & 0.0443 \\
    2442004 & 3972 & 8.86 & 1.031 & 0.0764 & 0.0911 \\
    ... & ... & ... & ... & ... & ... \\

    \bottomrule
  \end{tabular}
\end{table}

Several issues need to be addressed before we conduct further analysis.
First, while \emph{Kepler} observations cover 17 quarters, one or more quarters may lack data. 
Second, ref. \citep{2017ApJ...849...36Y} manually removed some flares flagged as contaminated. 
Consequently, an inferred waiting time longer than 90 days could be spurious.
To avoid artifacts in waiting time calculations, we adopt the following criteria: 
(1) we only consider targets with more than 17 flares; 
(2) if a waiting time corresponds to two flares separated by one or more quarters, it is excluded from the sample.
Finally, to ensure completeness in flare detection, we restrict our analysis to flares with energies exceeding $10^{31}$ erg following ref. \citep{2014ApJ...797..121H}. 

Flare waiting time may be affected by closely spaced, complex, or temporally clustered flares. In the catalog of ref. \cite{2017ApJ...849...36Y}, complex or multi-peaked flares are treated as a single flare rather than artificially splitting them. Meanwhile, in our sample, the shortest waiting time is roughly 0.1 day, meaning that two flares are separated by about 2.4 hours ($\sim$ 5 data points). Together, these measures effectively avoid too small waiting times. Then the median waiting time of each target is calculated and is marked as $t_{\rm{w, med}}$. 
The final results, along with the rotation periods ($P_{\rm{rot}}$) and effective temperatures ($T_{\rm{eff}}$) gathered from ref. \cite{2017ApJ...849...36Y}, are listed in Table \ref{tab:star}.
 
\section{Results and Discussion}
\label{sec:res}
\subsection{Comparisons to other activity proxies}

Flares are intensive energy release processes during magnetic reconnection. Consequently, flare parameters, including waiting time, morphology, frequency distribution, and flare rates, can serve as effective probes of stellar magnetic activity and dynamo. Among these parameters, flare rate is usually used to describe the mean activity level \citep{2017ApJ...849...36Y, 2020ApJ...905..107M} while waiting time serves as a critical diagnostic for distinguishing stochastic processes from intrinsic temporal correlations driven by magnetic energy release mechanisms \citep{2000ApJ...536L.109W, 2014ApJ...797..121H}. 

Combining our results with those of ref. \cite{2017ApJ...849...36Y} we plot $t_{\rm{w, med}}$ against flare rate (left panel of Figure \ref{picture_rate.fig}). To quantify the relation, we apply a uniform prior to the parameters. The likelihood is written as:
\begin{equation}
    \ln \mathcal{L}(\theta \mid x, y, \sigma_y) = -\frac{1}{2} \sum_{i=1}^{N} \left[ \frac{\left( y_i - (a x_i + b) \right)^2}{\sigma_{y,i}^2 + \sigma_{\mathrm{int}}^2} + \ln\left( \sigma_{y,i}^2 + \sigma_{\mathrm{int}}^2 \right) \right].
\end{equation}
Here $\sigma_{\rm{y,i }}$ is the error of flare rate, which is estimated as 1 / $\sqrt{N}$ and $N$ is the flare number. $\sigma_{\rm{int}}$ is the intrinsic scatter. The best-fit result is:
\begin{equation}
  \text{log}_{10}({\text{Rate}}) = (-0.99 \pm 0.01) \times \text{log}_{10}(t_{\rm{w, med}}) + (-0.32 \pm 0.01),  
\end{equation}
which is shown as red line. The slope is very close to $-$1, suggesting that the $t_{\rm{w, med}}$ can also trace the long-term stellar magnetic activity level.

\begin{figure}[H]
\centering
\includegraphics[width=0.45\textwidth]{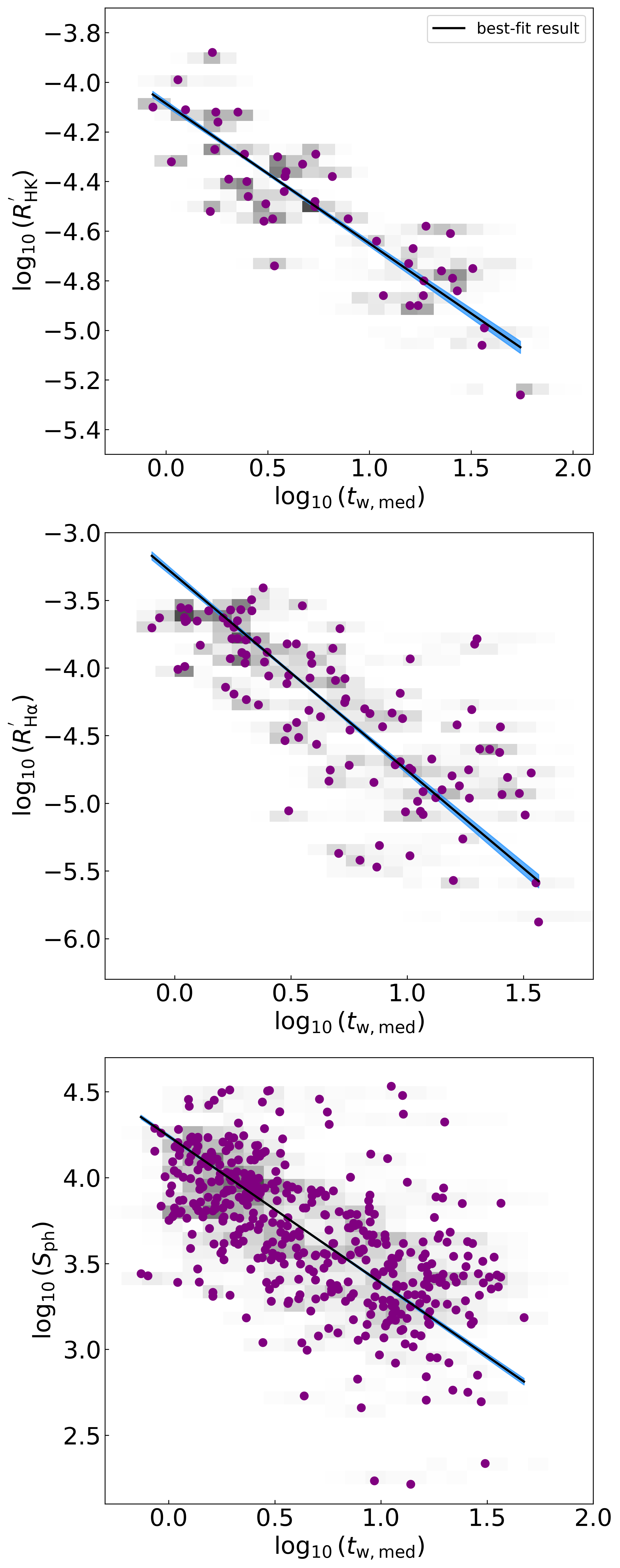}
\caption{Top panel: Comparison between $t_{\rm{w,med}}$ and $R_{\rm{HK}}^{'}$ (ref. \citep{2026ApJ..1000..208H}). Middle panel: Comparison between $t_{\rm{w,med}}$ and $R_{\rm{H\alpha}}^{'}$. Bottom panel: Comparison between $t_{\rm{w,med}}$ and $S_{\rm{ph}}$ (ref. \citep{2019ApJS..244...21S}). Black lines are best-fit results. Blue shaded areas indicate the 1$\sigma$ confidence boundaries obtained from the bootstrap distributions of the fitted lines.}
\label{picture.fig}
\end{figure}

Meanwhile, similar to previous studies, we propose that flare waiting times offer richer insights into underlying physical processes than flare rates alone. Waiting times are closely linked to the energy storage and release mechanisms, a relationship that remains actively investigated \citep{1976ApJS...30...85L, 1999ApJ...515..746O, 2010ApJ...710.1324W, 2014ApJ...797..121H}. Furthermore, $t_{\rm{w, med}}$ derived from different percentiles (10th, 90th, and overall) encode distinct information about stellar flares. For example, the right panel of Figure \ref{picture_rate.fig} shows that while $t_{\rm{w, med}}$ values for the 90th percentile and the full sample exhibit clear dependencies on flare rate, the $t_{\rm{w, med}}$ for the 10th percentile presents only a weak and scattered correlation. Thus, we propose that waiting times can probe not only average flare frequency but also temporal behavior. An exponential distribution implies independent Poisson events, while deviations suggest memory effects \citep{2026ApJ..1001..171P}. Moreover, the tail slope of the distribution may further provide observational constraints on dynamo process, which are not accessible from traditional indices like overall occurrence rates or normalized flare energy alone.

Different activity proxies could reveal magnetic activities in different layers of the stellar atmosphere \citep{1981ApJS...45..635V, 2017ARA&A..55..159L}. To test whether $t_{\rm{w, med}}$ correlates with magnetic proxies from other atmospheric layers, we compare it with three different proxies: $R_{\rm{HK}}^{'}$ and $R_{\rm{H\alpha}}^{'}$, which quantify the excess chromospheric emission in the \cahk lines and the $\rm{H\alpha}$ line, respectively;  $S_{\rm{ph}}$, which measures the standard deviation of the light curve amplitude (on a timescale of $5\times P_{\rm{rot}}$), reflecting spot-induced variability.
$R_{\rm{HK}}^{'}$ values are gathered from ref. \citep{2026ApJ..1000..208H}, while the photospheric activity proxy $S_{\rm{ph}}$ is taken from ref. \citep{2019ApJS..244...21S}. 
We calculate $R_{\rm{H\alpha}}^{'}$ following the procedures of ref. \citep{2017ApJ...834...85N, 2023ApJS..264...12H} (see Appendix \ref{sec:app} for more details). 

Obviously, $t_{\rm{w, med}}$ correlates well with all other activity proxies (Figure \ref{picture.fig}), further confirming that $t_{\rm{w, med}}$ is an effective magnetic activity proxy. Since flares are strongly correlated to magnetic fields, $t_{\rm{w, med}}$ should in principle be applicable to G- and K-type dwarfs as well. However, given that M dwarfs exhibit stronger magnetic activity and consequently present more flares, this study focuses exclusively on M dwarfs.

We then fit linear models to these relations using orthogonal distance regression (ODR) based on the ODR Python package from \emph{scipy}. To estimate uncertainties on the slope and intercept, we perform a non-parametric bootstrap.
First, for each target, we randomly draw one waiting time from its full set and repeat this process $\left\lceil 0.7 \times N \right\rceil$ times, where \emph{N} is the total number of waiting times for that target. Second, we compute the $t_{\rm{w, med}}$ from these realizations and fit the ODR model to the resampled data. Third, we repeat the entire procedure 10,000 times. The distribution of the derived $t_{\rm{w,med}}$ across these 10,000 realizations is shown as the gray shaded regions in Figure \ref{picture.fig}. Finally, we calculate the standard deviations of the 10,000 fitted slopes and intercepts and adopt them as the uncertainties in the fitted parameters.

\begin{figure}[H]
\centering
\includegraphics[width=0.42\textwidth]{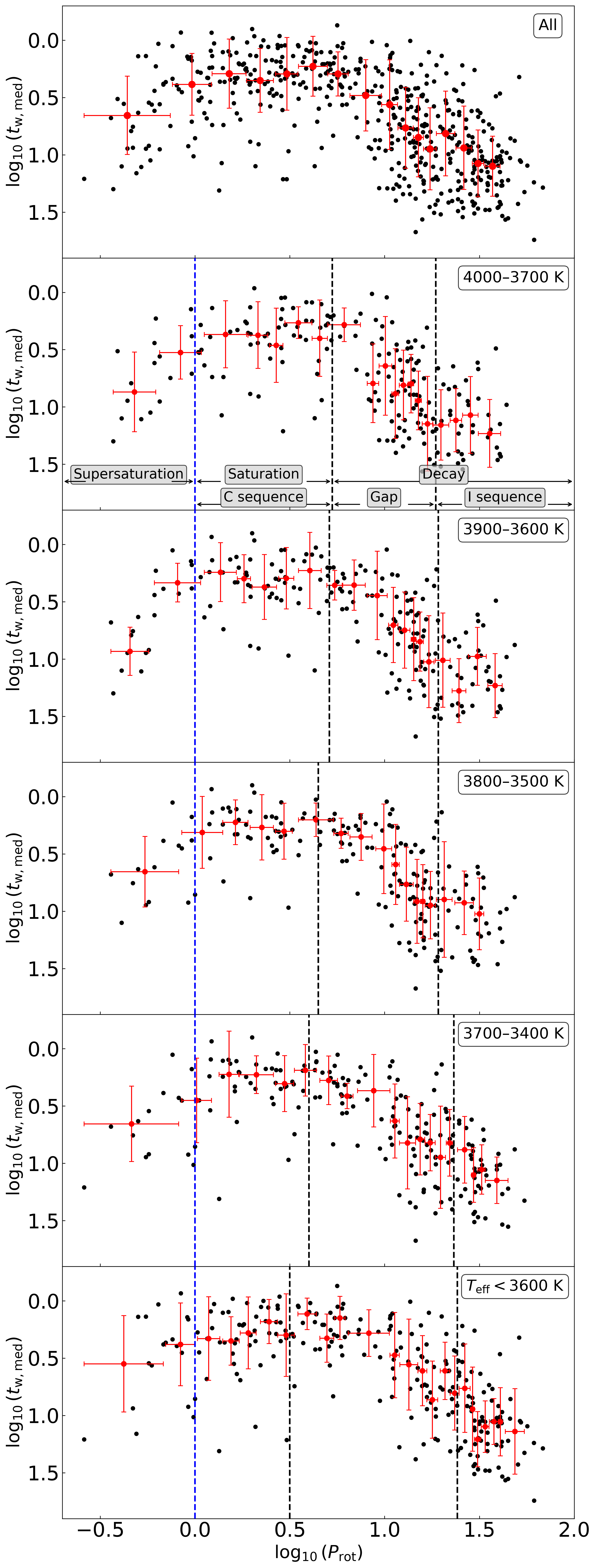}
\caption{Top panel: $t_{\rm{w, med}}$--$P_{\rm{rot}}$ relation of whole sample. Lower panels: $t_{\rm{w, med}}$--$P_{\rm{rot}}$ relations across different $T_{\rm{eff}}$ bins. Blue dashed lines mark the visually estimated boundary of the supersaturation region. Black dashed lines are from ref. \citep{2026ApJ..1000..208H}, which mark the transition points of different stages of gyrochronology.}
\label{picture2.fig}
\end{figure}

We derive the $t_{\rm{w, med}}-R_{\rm{HK}}^{'}$ relation, the $t_{\rm{w, med}}-R_{\rm{H\alpha}}^{'}$ relation, and the $t_{\rm{w, med}}-S_{\rm{ph}}$ relation as
\begin{equation}
    \log_{10}(R_{\rm{HK}}^{'}) = (-0.56 \pm 0.02) \times \log_{10}(t_{\rm{w, med}}) - (4.09 \pm 0.01),
\end{equation}

\begin{equation}
    \log_{10}(R_{\rm{H\alpha}}^{'}) = (-1.44 \pm 0.05) \times \log_{10}(t_{\rm{w, med}}) - (3.31 \pm 0.03),
\end{equation}
and
\begin{equation}
    \log_{10}(S_{\rm{ph}}) = (-0.85 \pm 0.02) \times \log_{10}(t_{\rm{w, med}}) + (4.24 \pm 0.01),
\end{equation}
respectively. 

\subsection{Activity--rotation relation}
Historically, the activity--rotation relation has been constructed using various activity proxies. However, the relation between $t_{\rm{w, med}}$ and $P_{\rm{rot}}$ is rarely investigated. This study is the first to directly link the $t_{\rm{w, med}}$ of flare events with the $P_{\rm{rot}}$. We also divide stars into different $T_{\rm{eff}}$ bins (Figure \ref{picture2.fig}) since stars with similar $T_{\rm{eff}}$ have comparable convective turnover timescales, which can remove the effect of stellar masses. 

We apply a binned median analysis to the overall $t_{\rm{w, med}}-P_{\rm{rot}}$ relation and the relation in each $T_{\rm{eff}}$ bin, requiring 30 targets or 10 targets in each bin, respectively. Each bin is represented by the median value of $t_{\rm{w, med}}$ and the central value of $P_{\rm{rot}}$. The error bars along the X-axis indicate the bin width, while those along the Y-axis show the standard deviation of $t_{\rm{w, med}}$ within each bin. 
Similar to the classical relation constructed using multiple proxies \citep{2011ApJ...743...48W, 2014ApJS..211...24M, 2017ApJ...834...85N}, the $t_{\rm{w, med}}$--$P_{\rm{rot}}$ relation mainly consists of two parts: the saturated region, where stellar activity remains unchanged, and the decay region, where the activity level becomes dependent on the rotation rate.

Notably, fine structures reported in previous literature are also evident. Recent studies proposed that the classical saturation and decay regions can be divided into three parts \citep{2025A&A...699A.251Y, 2026ApJ..1000..208H}, corresponding to the ``Convective'' sequence (``C" sequence), Gap, and ``Interface'' sequence (``I" sequence) in gyrochronology \citep{2003ApJ...586..464B}. 
Generally, the saturation region matches the ``C'' sequence, while the decay region corresponds to the Gap and ``I'' sequence. In lower panels of Figure \ref{picture2.fig}, we plot knee points (with black dashed lines) from ref. \citep{2026ApJ..1000..208H}, marking the transition points of different stages of gyrochronology: log$_{10}(P_{\rm rot}) \approx 0.5$ for the transition from the ``C'' sequence to the Gap and log$_{10}(P_{\rm rot}) \approx 1.3$ for the transition from the Gap to the ``I" sequence. Note that the transition points are mass-dependent.
The slope of the decay rate shows a clear variation from the Gap to the ``I" sequence: the $t_{\rm{w, med}}-P_{\rm{rot}}$ relation gradually flattens, indicating a lower decay rate.

Intriguingly, we also notice a region ($P_{\rm{rot}} < 1$ day) where $t_{\rm{w, med}}$ values  increase significantly with faster rotation. 
This aligns with the supersaturation phenomenon, which was first identified in X-ray observations by ref. \citep{1996A&A...305..785R} and later confirmed using other activity indicators \citep{2001A&A...370..157S, 2011ApJ...743...48W, 2024ApJS..273....8H}. In this region, fast-rotating stars exhibit a lower activity level. The supersaturation phenomenon observed in the $t_{\rm{w, med}}-P_{\rm{rot}}$ relation is more notable than that found in other activity proxies. 
In addition, these fine structures of the $t_{\rm{w, med}}-P_{\rm{rot}}$ relation are quite similar to those of the X-ray activity--rotation relation, which also exhibits a four-part pattern \citep{2024ApJS..273....8H}. 

\begin{figure}[H]
\centering
\includegraphics[width=\columnwidth]{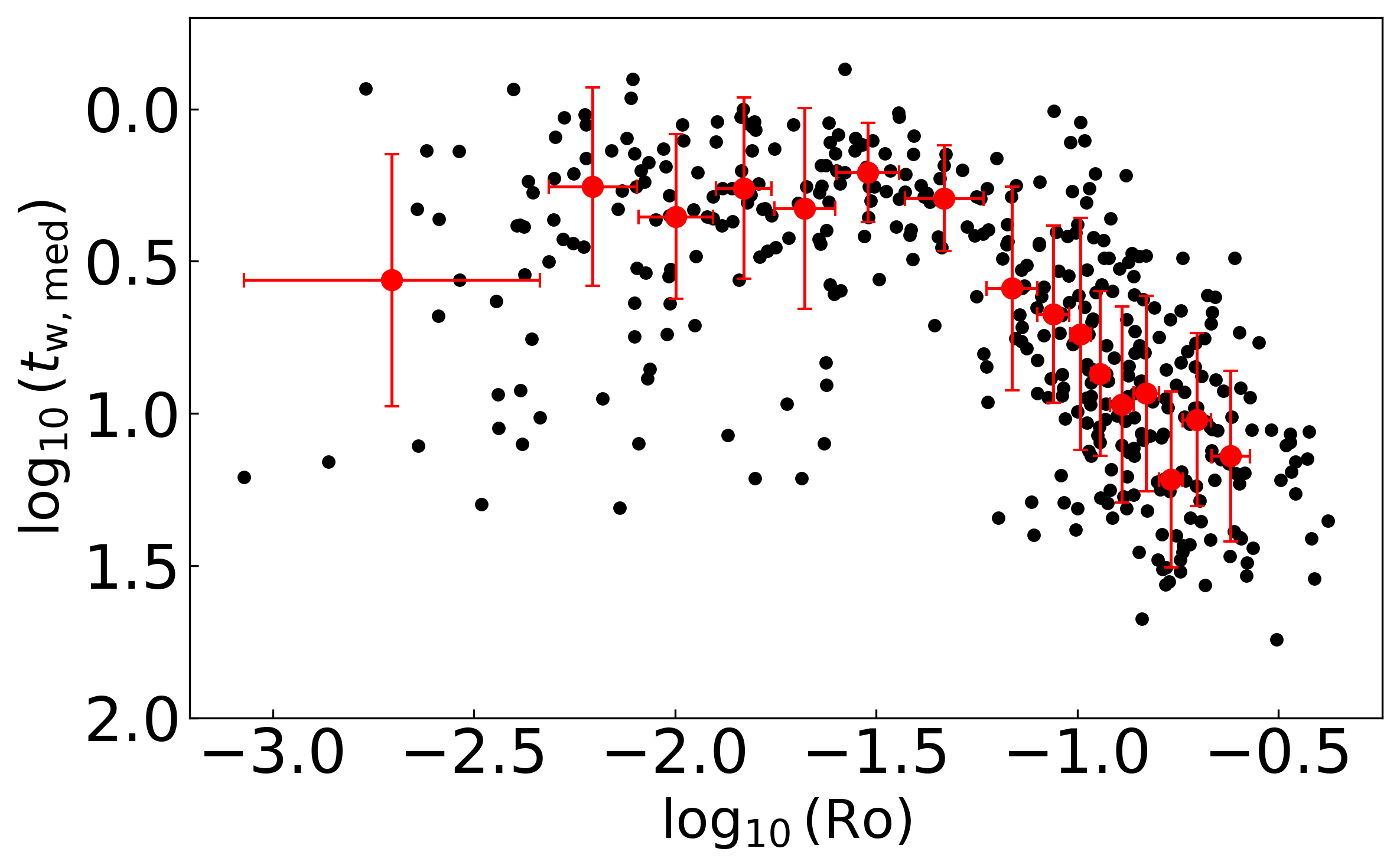}
\caption{$t_{\rm{w, med}}-$Ro relation.}
\label{pictureB.fig}
\end{figure}

Meanwhile, many studies prefer using the Rossby number (Ro = $P_{\rm{rot}}/\tau_{c}$, where $\tau_{c}$ is convective turnover time) over $P_{\rm{rot}}$ when constructing the activity--rotation relation, as Ro eliminates stellar mass effects. To calculate the Rossby number, we first collect stellar parameters, including $T_{\rm{eff}}$, log$g$ and [Fe/H], from the StarHorse catalog \citep{2022A&A...658A..91A} and use the LAMOST Data Release 12 (DR12) catalog \citep{2015RAA....15.1095L} as a supplement. Then we use the Yale-Potsdam Stellar Isochrones (YAPSI) \citep{2017ApJ...838..161S} containing solar calibration with metallicity grids of 0.3, 0.0, $-$0.5, $-$1.0, $-$1.5 to calculate the $\tau_{\rm{c}}$. For each available metallicity grid, a two-dimensional interpolator is built over the $T_{\rm{eff}} - \text{log}g$ plane. For a given star, if its [Fe/H] lies within the range of the grid slices, the $\tau_{\rm{c}}$ values from the two neighbouring slices are first obtained via two-dimensional interpolation, then linearly interpolated along the [Fe/H] dimension. Stars with [Fe/H] above or below the highest or lowest available grid are handled by a direct two-dimensional interpolation using that highest-metallicity or lowest-metallicity grid. We plot the $t_{\rm{w, med}}-$Ro relation in Figure \ref{pictureB.fig}, which is quite similar to the $t_{\rm{w, med}}-P_{\rm{rot}}$ relation.

\begin{figure*}
\centering
\includegraphics[width=0.95\textwidth]{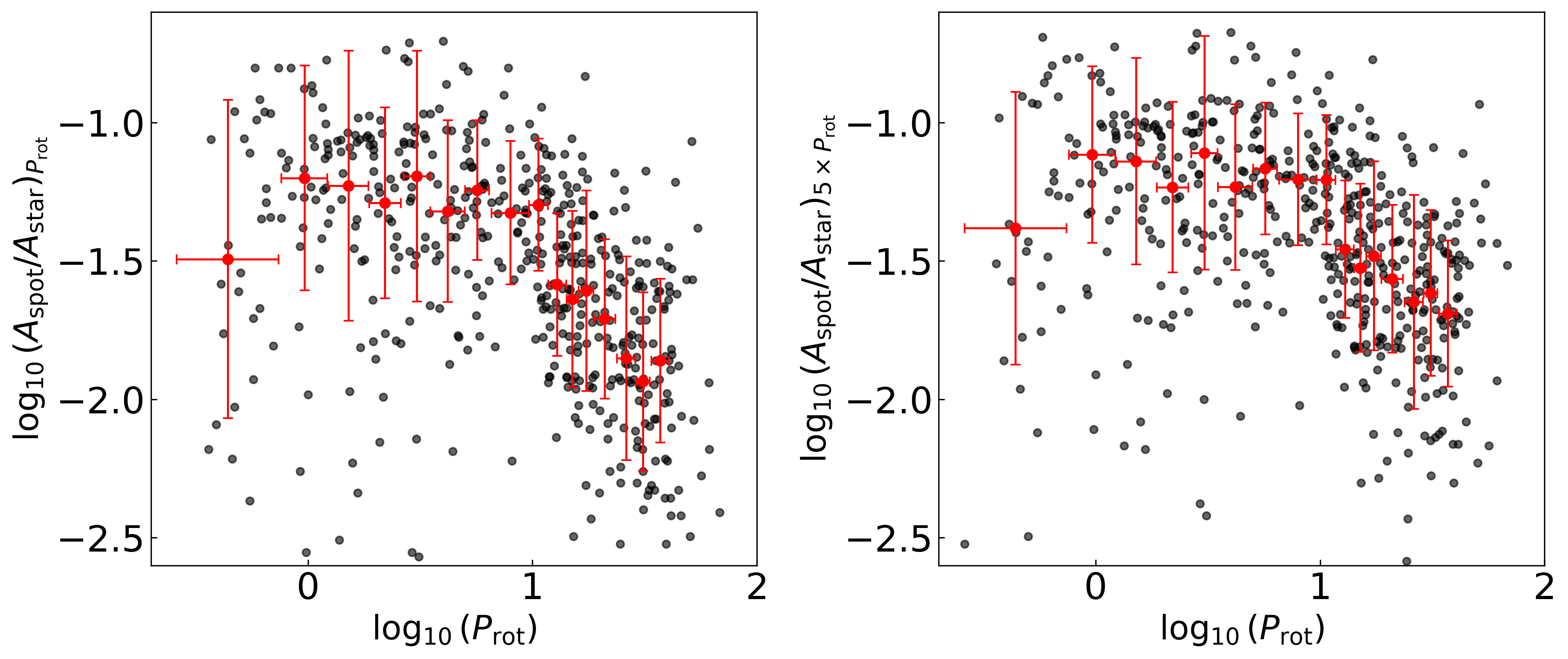}
\caption{Relations between the filling factor and rotation period. Left panel shows the result of $(\frac{A_{\rm{spot}}}{A_{\rm{star}}})_{{P_{\rm{rot}}}}$ while the right panel shows the result of $(\frac{A_{\rm{spot}}}{A_{\rm{star}}})_{{5\times P_{\rm{rot}}}}$. }
\label{picture3.fig}
\end{figure*}

\subsection{Explanations for Supersaturation}

Two main explanations have been proposed for the supersaturation phenomenon. One involves a reduced X-ray emitting volume due to centrifugal stripping, which strips away part of the corona and leads to open magnetic field configurations that appear as dark regions in X-rays \citep{2000MNRAS.318.1217J, 2004A&A...414L...5J}. However, this scenario requires large, low-density loops, which have not been supported by prior observations \citep{2006ApJ...650.1119H}. Meanwhile, the supersaturation effect has also been observed in chromospheric activity proxies \citep{2024ApJ...976..243D, 2025MNRAS.542.2431L}, indicating that centrifugal stripping is unlikely to be the underlying cause.

The other possible explanation was proposed by ref. \citep{2001A&A...370..157S}, which suggested that extremely fast rotating stars tend to have lower filling factors of active regions. According to the Zeipel theorem \citep{1924MNRAS..84..665V}, these stars will present a higher fraction of flux at the poles. Such flux imbalance can drive strong convective updrafts in the outer convective envelope \citep{2001A&A...370..157S}, sweeping interior magnetic flux tubes poleward before they emerge at the surface. As a result, active regions will gather at the poles, resulting in a reduced filling factor. 

To test this scenario, we investigate the relation between the filling factor and $P_{\rm{rot}}$. Following ref. \citep{2017PASJ...69...41M}, filling factor is defined as:
\begin{equation}
    \frac{A_{\rm{spot}}}{A_{\rm{star}}} = \frac{\Delta F}{F} \frac{T_{\rm{star}}^{4}}{T_{\rm{star}}^{4} - T_{\rm{spot}}^{4}}.
\end{equation}
$T_{\rm{spot}}$ is the temperature of starspot calculated following ref. \citep{2005LRSP....2....8B}:
\begin{equation}
    T_{\rm{spot}} = T_{\rm{star}} - 3.58\times 10^{-5}\times T_{\rm{star}}^{2} - 0.249 \times T_{\rm{star}} + 808.
\end{equation}
$T_{\rm{star}}$ is the stellar effective temperature. $\frac{\Delta F}{F}$ is the normalized amplitude of light curves modulated by stellar rotation. 

Considering that frequent flares will influence the measurement of amplitudes of rotational modulation, we utilize the flare-cleaned light curve, i.e., the baseline after $\sigma$-clipping, provided by ref. \citep{2017ApJ...849...36Y} to calculate $\frac{\Delta F} {F}$, which is usually defined over some representative timescales. In this work, two widely adopted timescales are considered: (1) Following ref. \citep{2017PASJ...69...41M,2014ApJS..211...24M}, we divide the normalized light curve of each target into segments of length $1 \times P_{\rm{rot}}$ and compute the maximum amplitude within each segment. The median of these amplitudes is then adopted as the final $\frac{\Delta F}{F}$. The corresponding filling factor is represented as $(\frac{A_{\rm{spot}}}{A_{\rm{star}}})_{P_{\rm{rot}}}$. (2) Motivated by ref. \citep{2014A&A...562A.124M}, who reported that the standard deviation of a light curve over a timescale of $5 \times P_{\rm{rot}}$ provides a robust proxy for photospheric activity while minimizing contamination from other effects such as stellar cycles, we apply an analogous procedure to the flare-cleaned light curves but using segments with $5 \times P_{\rm{rot}}$ length. The corresponding filling factor is marked as $(\frac{A_{\rm{spot}}}{A_{\rm{star}}})_{{5\times P_{\rm{rot}}}}$. 

Figure \ref{picture3.fig} shows filling factor--rotation relations constructed with $(\frac{A_{\rm{spot}}}{A_{\rm{star}}})_{P_{\rm{rot}}}$ and $(\frac{A_{\rm{spot}}}{A_{\rm{star}}})_{{5\times P_{\rm{rot}}}}$, which are similar to the $t_{\rm{w, med}}$--$P_{\rm{rot}}$ relation. It is obvious that for both the $(\frac{A_{\rm{spot}}}{A_{\rm{star}}})_{P_{\rm{rot}}}-P_{\rm{rot}}$ relation and $(\frac{A_{\rm{spot}}}{A_{\rm{star}}})_{5\times P_{\rm{rot}}}-P_{\rm{rot}}$ relation, in the regime where $P_{\rm{rot}} < 1$ day, the filling factor clearly decreases with faster rotation. This suggests that supersaturation may be driven by reduced filling factors, thus favoring the poleward migration scenario.

Other possible explanations also exist. For example, ref. \citep{2026SCPMA..6989511W} recently investigated stellar magnetic dynamos in non-spherical M dwarfs within cataclysmic variables and found that strong distortion weakens thermal convection and, consequently, differential rotation.
These effects can in turn reduce magnetic activity.
Fast-rotating stars are also non‑spherical, and the resulting suppression of thermal convection may similarly account for the supersaturation regime. Addressing this hypothesis requires further observational and simulation efforts.

\section{Summary}
\label{sec:sum}
In this work, utilizing flaring M dwarfs, we investigate statistical properties of median flare waiting times. $t_{\rm{w, med}}$ exhibits a strong correlation with flare rate, making it a reliable tracer of long-term stellar activity. Meanwhile, we find that $t_{\rm{w, med}}$ is inversely correlated to other activity proxies, further suggesting that $t_{\rm{w, med}}$ can serve as a simple but effective activity proxy. In addition, for the first time we construct the $t_{\rm{w, med}}$--rotation relation for M dwarfs.
Similar to X-ray observations, we propose that the $t_{\rm{w, med}}-P_{\rm{rot}}$ relation consists of four parts:
a supersaturation region where $t_{\rm{w, med}}$ increases with faster rotation;
a saturation region where $t_{\rm{w, med}}$ keeps constant, corresponding to the ``C" sequence in the gyrochronology;
a fast decay region and a slow decay region, corresponding to the Gap and the ``I'' sequence, respectively.
Based on the relation between filling factor and stellar rotation period, we favor active region migration over coronal stripping as the explanation for supersaturation.

\vspace{0.6cm}

{\footnotesize
This work made use of the data from LAMOST (Large Sky Area Multi-Object Fiber Spectroscopic Telescope, also known as the Guoshoujing Telescope) (https://cstr.cn/31118.02.LAMOST). LAMOST is a Chinese national mega-science facility, operated by National Astronomical Observatories, Chinese Academy of Sciences.
This work was supported by National Natural Science Foundation of China (NSFC) under grant Nos. 12588202/12273057/11833002/12090042, the National Key Research and Development Program of China (NKRDPC) under grant number 2023YFA1607901, the Strategic Priority Program of the Chinese Academy of Sciences under grant number XDB1160302, and science research grants from the China Manned Space Project. J.F.L acknowledges the support from the New Cornerstone Science Foundation through the New Cornerstone Investigator Program and the XPLORER PRIZE. C.J.Z acknowledges the support from China Postdoctoral Science Foundation under grant number 2025M783229.
}

\bibliographystyle{scibull}
\bibliography{refer}

\newpage


\renewcommand{\thesection}{Supplementary Materials}
\begin{appendices}
\section*{APPENDIX}                         
\setcounter{table}{0}   
\setcounter{figure}{0}
\setcounter{equation}{0}
\renewcommand{\thetable}{A\arabic{table}}
\renewcommand{\thefigure}{A\arabic{figure}}
\renewcommand{\theequation}{A\arabic{equation}}
\section{$\rm{H{\alpha}}$ emissions and $R_{\rm{H\alpha}}^{'}$} 
\label{sec:app}

\begin{figure}[H]
\centering
\includegraphics[width=\columnwidth]{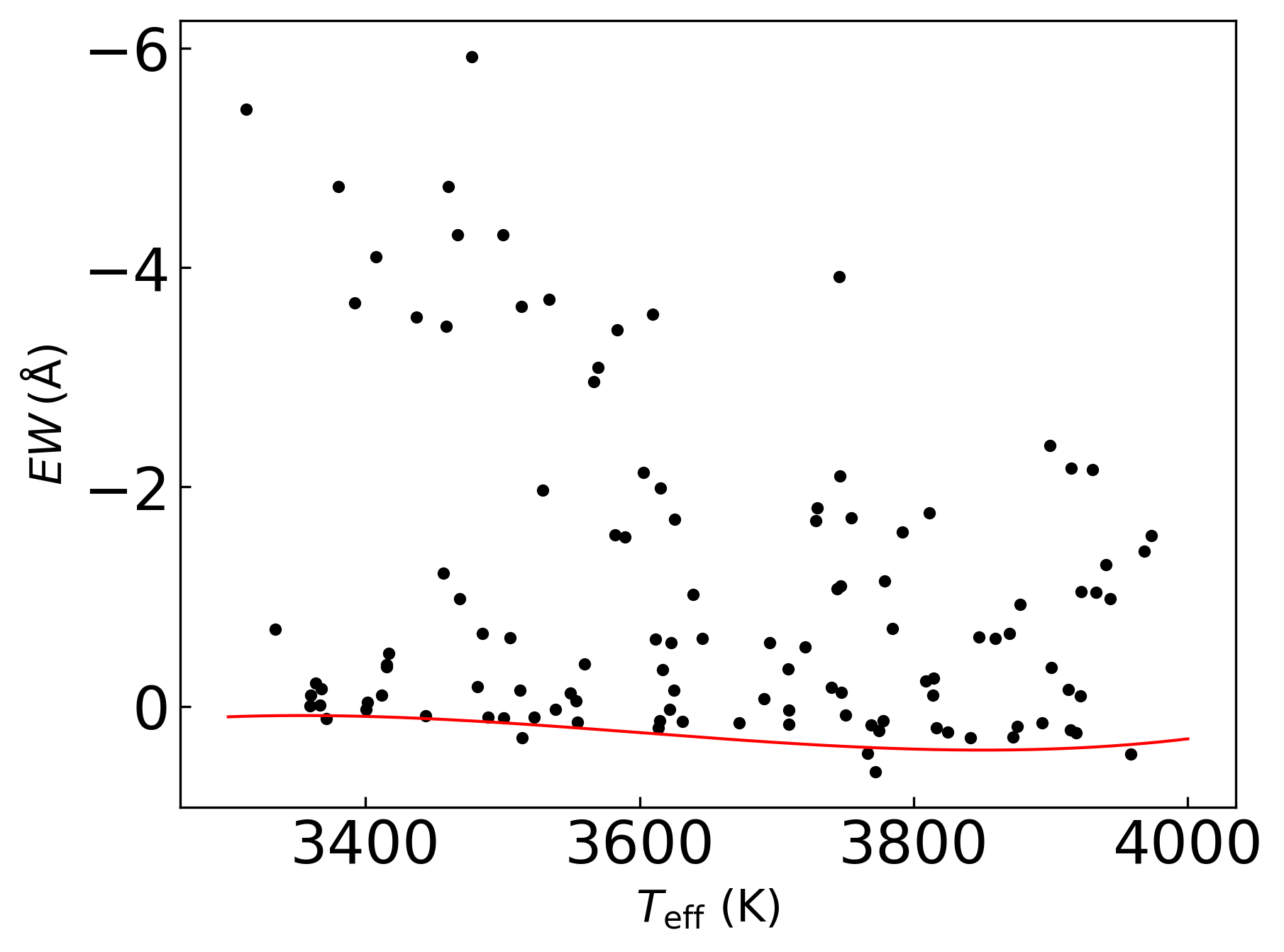}
\caption{$EW-T_{\rm{eff}}$ diagram. Red line represents the baseline.}
\label{pictureA.fig}
\end{figure}

To quantify the strength of $\rm{H\alpha}$ emission, we follow the procedures of ref. \citep{2004PASP..116.1105W, 2017ApJ...834...85N, 2023ApJS..264...12H}. First, we gather LAMOST low-resolution spectra with signal-to-noise ratio larger than 20 and $T_{\rm{eff}}$ of M dwarfs from LAMOST DR12 \cite{2015RAA....15.1095L}. All the spectra are radial velocity corrected using the redshift provided by the LAMOST DR12. Then the spectra are normalized using the Python package from ref. \citep{2020ApJS..246....9Z}, which integrated many aspects of dealing with LAMOST spectra, including normalization and radial velocity measurements. Then we calculate equivalent widths of $\rm{H\alpha}$ lines as
\begin{equation}
    EW = \int\frac{F_{\rm{c}} - F_{\rm{\lambda}}}{F_{\rm{c}}}d\lambda.
\end{equation}
Here $F_{\rm{c}}$ is the median value of the pseudocontinua within 10 \AA \, widths at both sides of line wings. The integration interval is set to be 10 \AA \, centered at 6564.6 \AA. Obviously H$\rm{\alpha}$ emissions will have negative $EW$s.

The H$\alpha$ line contains contributions from photosphere, which need to be subtracted to derive pure chromospheric activity level. One of the widely used methods is to subtract a baseline in the $EW-T_{\rm{eff}}$ diagram \citep{2017ApJ...834...85N}. To establish the baseline of the $EW-T_{\rm{eff}}$ relation, we employ a binned percentile fitting approach. The data are divided into 100 K bins in $T_{\rm{eff}}$. Each bin is marked using the median value of $T_{\rm{eff}}$. Then we compute the median value of the 98th percentile of $EW$s and fit the $EW-T_{\rm{eff}}$ relation using a third-order polynomial. During the fitting process, we apply one iteration of 3$\sigma$-clipping to reject outliers and obtain an optimal baseline (Figure \ref{pictureA.fig}). Then the chromospheric contribution to $EW$ is defined as:
\begin{equation}
    EW^{'} = EW - EW_{\rm{base}}.
\end{equation}
Finally, we convert the $EW^{'}$ to fractional $\rm{H\alpha}$ luminosity ($R_{\rm{H\alpha}}^{'}$) following
\begin{equation}
    R_{\rm{H\alpha}^{'}} = \chi \times EW^{'},
\end{equation}
where $\chi$ is defined as
\begin{equation}
    \chi = \frac{f_{\rm{6564}}}{\sigma T^{4}}.
\end{equation}
Here $f_{\rm{6564}}$ is the continuum flux at a wavelength of 6564 $\AA$, which was estimated using PHOENIX synthetic spectra \citep{2013A&A...553A...6H} corresponding to the nearest LAMOST stellar parameters of our M dwarf sample.

\end{appendices}

\end{multicols}
\end{CJK*}
\end{document}